\documentclass[12pt]{article}
\usepackage{epsf}
\def\be{\begin{equation}}
\def\ee{\end{equation}}
\def\bea{\begin{eqnarray}}
\def\eea{\end{eqnarray}}

\def\gsim{\raisebox{-0.6ex}{$\stackrel{\textstyle >}{\sim}$}}

\begin{document}
\begin{titlepage}
\begin{center}
{\Large \bf William I. Fine Theoretical Physics Institute \\
University of Minnesota \\}
\end{center}
\vspace{0.2in}
\begin{flushright}
FTPI-MINN-19/02 \\
UMN-TH-3811/19 \\
January 2019 \\
\end{flushright}
\vspace{0.3in}
\begin{center}
{\Large \bf Strange hadrocharmonium.
\\}
\vspace{0.2in}
{\bf  M.B. Voloshin  \\ }
William I. Fine Theoretical Physics Institute, University of
Minnesota,\\ Minneapolis, MN 55455, USA \\
School of Physics and Astronomy, University of Minnesota, Minneapolis, MN 55455, USA \\ and \\
Institute of Theoretical and Experimental Physics, Moscow, 117218, Russia
\\[0.2in]

\end{center}

\vspace{0.2in}

\begin{abstract}
It has been recently suggested that the charged charmoniumlike resonances $Z_c(4100)$ and  $Z_c(4200)$ are two states of hadrocharmonium, related by the charm quark spin symmetry in the same way as the lowest charmonium states $\eta_c$ and $J/\psi$. It is pointed out here that in this picture one might expect existence of their somewhat heavier strange counterparts, $Z_{cs}$, decaying to $\eta_c K$ and $J/\psi K$. Some expected properties of such charmoniumlike strange resonances are discussed that set benchmarks for their search in the decays of the strange $B_s$ mesons.
  \end{abstract}
\end{titlepage}

Numerous new resonances recently uncovered near the open charm and open bottom thresholds, the so-called XYZ states, apparently do not fit in the standard quark-antiquark template and contain light constituents in addition to a heavy quark-antiquark pair. (Recent reviews of the data and of the theoretical approaches can be found in Refs.~\cite{dsz,Guo17,Ali17}.) It becomes clear that the internal dynamics of these essentially multi-body systems is likely very much different in different states~\cite{mvch}. In particular, some of these exotic resonances, with mass very near a threshold for a heavy meson pair appear to display properties characteristic for loosely correlated threshold state, a molecule~\cite{vo}, made of the meson pair, with such picture likely applicable to the $Z_b(10610)$ and $Z_b(10650)$ bottomoniumlike resonances~\cite{bellez}, and to the $X(3872)$~\cite{bellex}, $Z_c(3900)$~\cite{besz39} and $Z_c(4020)$\cite{besz40} in the charmoniumlike sector. Another type of states is apparently presented by those that are not especially close to any heavy meson pair threshold and tend to decay into a particular state of quarkonium and one or more light mesons. Combined with the observation that the latter resonances do not overwhelmingly decay into pairs of heavy mesons, this has led to the `hadroquarconium' picture~\cite{mvch,dv} of such resonances, where a compact state of quarkonium is embedded into an excited light mesons by a QCD analog of the van der Waals force. The observed decays into quarkonium and light meson(s) are then due to the de-excitation of the light degrees of freedom. The hadrocharmonium picture, originally suggested for explaining the properties of the $J/\psi \pi \pi$ resonance $Y(4260)$ (lately ``shifted'' in mass down to about 4220\,MeV), has been recently invoked~\cite{mv18} for description of the charged charmoniumlike resonances  $Z_c(4100)$ and  $Z_c(4200)$ observed respectively in the decay channels $\eta_c \pi$~\cite{lhcbzc} and $J/\psi \pi$~\cite{bellezc}~\footnote{Some alternative models of the $Z_c(4100)$ resonance can be found in Refs.~\cite{zhao,cd,adkmnnz,saa}. Here however we use only the hadrocharmonium interpretation~\cite{mv18} of this state as well as of $Z_c(4200)$.}. 

It can be noted that some of the observed XYZ resonances have a nontrivial light flavor structure and come in isotopic triplets with charged components as well the neutral ones. However as of yet no states of this type with open strangeness have been observed. The (non)existence of molecular strange threshold systems could be closely related to the  problem of the forces between heavy mesons that give rise to the threshold singularities, in particular the significance of the pion exchange. The long-range interaction mediated by the pions has been much discussed in connection with the molecular states (see e.g. in Refs.~\cite{mp,nv,mv16,wbfhnw}). This interaction however is impossible between a non-strange and strange heavy meson, and the lightest exchanged meson is $\eta$, generally resulting in a somewhat different dynamics~\cite{kr}. Thus it is not likely that a straightforward application of the flavor SU(3) symmetry to the threshold states can be justified.

The situation looks quite different for the hadrocharmonium type systems where the flavor SU(3) may be applicable. Indeed, in these systems the compact heavy quarkonium interacts with the `hosting' light-quark resonance by exchange of gluons, rather than of quarks.  Thus an application of the SU(3) symmetry is the standard one for the light-quark excited states. As is well known the symmetry works reasonably well for such objects with thye main effect being that the strange states are heavier than their non-strange partners by $\Delta \approx (150 - 200)$\,MeV due to larger mass of the strange quark. Based on this argument one can expect that non-strange light-flavor non-singlet hadrocharmonium resonances should have strange analogs that are heavier by about the same amount $\Delta$ and have similar large widths in the ballpark of (100 - 400)\,MeV. 

In the model where the $Z_c(4100)$ and $Z_c(4200)$ are states of hadrocharmonium with respectively $\eta_c$ and $J/\psi$ bound in $S$ wave to the same light quark excitation with the quantum numbers of a pion, one thus can expect existence of two similar strange states with the pion excitation being replaced by one with quantum numbers of a Kaon. It is known~\cite{pdg} that the observed among light mesons Kaon excitation $K(1460)$ is by $\sim 160$\,MeV heavier than its non strange analog $\pi(1300)$.  Thus the strange hadrocharmonium resonances with quantum numbers $J^P=0^+$ and $1^+$ can be expected with the mass around 4250\,MeV and 4350\,MeV, and, for concreteness, these will be referred here as $Z_{cs}(4250)$ and $Z_{cs}(4350)$. Certainly, the specific values of the masses of both the known non-strange $Z_c(4100)$ and $Z_c(4200)$ as well as of the hypothetical strange states are subject to a considerable uncertainty due to their large widths. (For illustration: the PDG tables quote a $\pm 100$\,MeV uncertainty for the mass of the $\pi(1300)$ resonance.) 

\begin{figure}[ht]
\begin{center}
 \leavevmode
    \epsfxsize=16.5cm
    \epsfbox{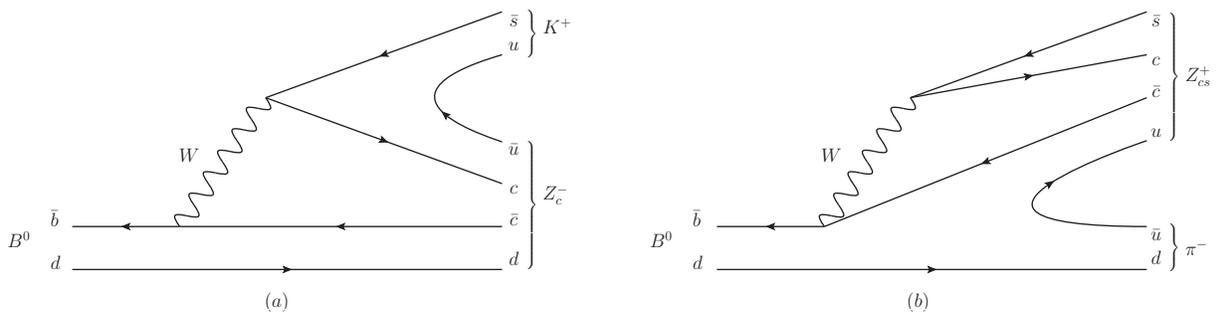}
    \caption{The quark graphs for the decays $B^0 \to Z_c^- K^+$ (a) and $B^0 \to Z_{c}^+ \pi^-$ (b).  }
\end{center}
\end{figure}

\begin{figure}[ht]
\begin{center}
 \leavevmode
    \epsfxsize=16.5cm
    \epsfbox{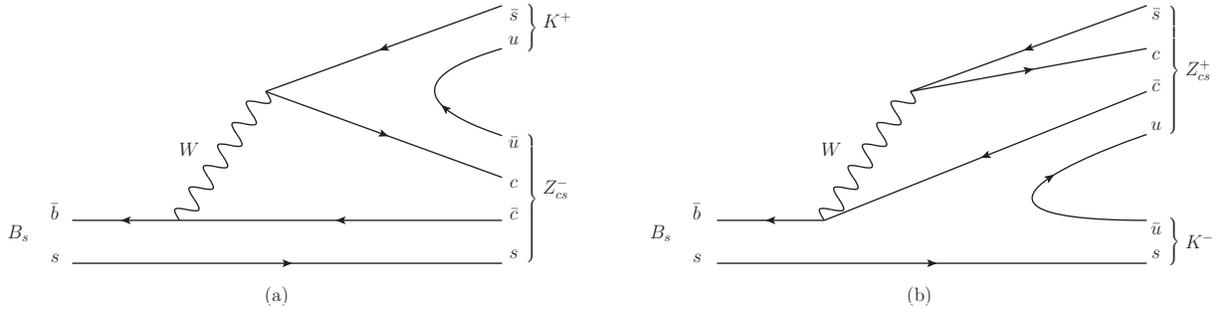}
    \caption{The quark graphs for the decays of strange $B_s$ meson, $B_s \to Z_{cs}^- K^+$ (a) and $B_s \to Z_{cs}^+ K^-$ (b).  }
\end{center}
\end{figure}

Clearly, it should be expected that a significant, if not dominant, decay modes of the suggested strange resonances should be $Z_{cs}(4250) \to \eta_c K$ and $Z_{cs}(4350) \to J/\psi K$, and they can be sought for in the decays of non-strange and strange $B$ mesons, e.g. in the processes $B^0 \to Z_{cs}^+ \pi^- $,  $B_s \to Z_{cs}^+ K^-$ or $B_s \to Z_{cs}^- K^+$. Out of these processes only the latter decay is a direct analog of the processes where the $Z_c(4100)$ and $Z_c(4200)$ resonances were observed~\cite{lhcbzc,bellezc}: $B^0 \to Z_c(4100) K^+ \to \eta_c \pi^- K^+$ and $B^0 \to Z_c(4200) \to J/\psi \pi^- K^+$so that an approximate prediction for the rate can be deduced. The types of the $B^0$ meson decays are shown in Fig.~1 and for the strange $B_s$ mesons the relevant types of processes are shown in Fig.~2. (We use generic notation $Z_c$ and $Z_{cs}$ in the figures and in the text, where the discussion is the same for the $J^P=0^+$ and $1^+$ resonances.) The graph in Fig.~1a is described in Ref.~\cite{lhcbzc}, and it is clear from it that the $Z_c$ resonance gets the spectator quark from the parent $B$ meson. For this reason the decay of a non-strange $B$ meson into a final state with a strange $Z_{cs}^+$ resonance, $B^0 \to Z_{cs}^+ \pi^-$ would require absorbtion of the $\bar s$ antiquark emerging from the decay and would thus not be similar to the observed process of Fig.~1a. A similar to the observed production of non-strange $Z_c$ resonances  would rather be the decay of the strange $B_s$ meson that contains spectator $s$ quark as shown in Fig.~2a. In the approximation where the OZI suppressed $s \bar s$ annihilation is neglected, the amplitudes of the processes in Fig.~1 and Fig.~2 are simply related by the SU(3) flavor symmetry, implying that the processes in figures 1a and 2a are described by the same amplitude $A$, while those in the figures 1b and 2b are given by a separate common amplitude $B$. Neglecting the [SU(3) breaking] kinematical differences between the processes, one can write the rates of the discussed decays in terms of the amplitudes $A$ and $B$ as
\be
\Gamma(B^0  \to Z_c^- K^+) = |A|^2~, ~~\Gamma(B^0 \to Z_{cs}^+ \pi^-) = |B|^2
\label{b0dec}
\ee
and 
\be
\Gamma(B_s  \to Z_{cs}^- K^+) = |A|^2~, ~~\Gamma(B_s \to Z_{cs}^+ K^-) = |B|^2~.
\label{bsdec}
\ee
It should be noted that the latter relations are written for the flavor-specific state $B_s$, where the discussed two amplitudes describe different processes and there is no interference between them. For the mass eigenstates of the $B_s - \bar B_s$ mixing such interference arises with a relative phase determined by presently unknown dynamics. In the averaged rate however the interference disappears, modulo small effects of the lifetime difference, and one can use the expression for the branching fraction of an `average' $B_s$ meson as
\be
\Gamma[B_s \to Z_{cs}^- K^+] = {1 \over 2} \, \left ( |A|^2 + |B|^2 \right )={1 \over 2} \,\left [\Gamma(B^0  \to Z_c^- K^+)+  \Gamma(B^0 \to Z_{cs}^+ \pi^-) \right ]~.
\label{bsav}
\ee

Only the decays with the non-strange $Z_c$ resonances were observed so far with the branching fractions ${\cal B}[B^0 \to Z_c^-(4100) K^+] \sim 1.9 \times 10^{-5}$~\cite{lhcbzc} and ${\cal B}[B^0 \to Z_c^-(4200) K^+] \sim 2.2 \times 10^{-5}$~\cite{bellezc} [Only the central values are quoted here. The (quite substantial) experimental errors from various sources can be found in the original papers.] The relation (\ref{bsav}) thus sets a benchmark for the required sensitivity of a search for either of the discussed strange hadrocharmonium resonances as ${\cal B}(B_s \to Z_{cs}^- K^+) ~ \gsim ~ 1 \times 10^{-5}$. This value corresponds to few percent of the measured branching fraction for the known~\cite{lhcbpsikk,bellepsikk} decay $B_s \to J/\psi K^+ K^-$, similarly to the contribution of  non-strange $Z_c$ resonances to decays into charmonium plus a $K \pi$ pair,  . It appears however that the existing data on the $B_s$ decays are inconclusive  with regards to existence of the discussed $Z_{cs}$ resonance. Hopefully, a search for strange hadrocharmonium in the channel $J/\psi K$, as well as a search for $\eta_c K$ resonance in  $B_c \to \eta_c K^+ K^-$, can become possible in future data with higher statistics.

The amplitude $B$ in Eqs.~({\ref{b0dec}) and (\ref{bsdec}) is currently unknown. Thus it is impossible at present to estimate the necessary sensitivity of search for the suggested resonances in decays of non-strange $B$ mesons.

As is already mentioned, the dominant decays of the suggested strange resonances are likely to be $Z_{cs}(4250) \to \eta_c K$ and $Z_{cs}(4350) \to J/\psi K$. In addition one can expect sub-dominant decay modes into charmonium and light meson(s) that should be characteristic for hadrocharmonium, as discussed~\cite{mv18} for the non-strange $Z_c$ resonances. In particular, the decays with a breaking of the charm quark spin symmetry, $Z_{cs}(4250) \to J/\psi K^*(890)$ and $Z_{cs}(4350) \to \eta_c K^*(890)$ can be expected with branching fraction from several percent to a few tens percent and the widths related as~\cite{mv18}:
\be
\Gamma[Z_{cs}(4250) \to J/\psi K^*(890)] \approx 3 \, \Gamma[Z_{cs}(4350) \to \eta_c K^*(890)]~.
\label{kstar}
\ee
Furthermore there should be decays (at a suppressed rate) to the radially excited charmonium,
\be
 \Gamma[Z_{cs}(4250) \to \eta_c(2S) K] \approx  \Gamma[Z_{cs}(4350) \to \psi(2S) K]~,
\label{2sr}
\ee
and into the $P$ wave excitations (also requiring the spin symmetry breaking):
\be
{\Gamma[Z_{cs}(4350) \to h_c K]   \over \Gamma[Z_{cs}(4250) \to \chi_{c1} K]} \approx \left ( {p_2 \over p_1} \right )^3 \approx 1.7~,
\label{rp21}
\ee
where $p_2$ and $p_1$ are the values of the momentum in the two processes. 

To summarize. If the observed exotic resonances $Z_c(4100)$ and $Z_c(4200)$ are interpreted as states of hadrocharmonium related by the heavy quark spin symmetry, it would be natural to expect, based on the flavor SU(3) symmetry, existence of their strange partners, $Z_{cs}$, with a slightly heavier mass and similarly large widths. The estimated rate of production of these resonances in the decays of the strange $B_s$ meson, $B_s \to Z_{cs} K$, possibly makes feasible a search for these resonances in future data. The rate for the decays $B \to Z_{cs} \pi$ cannot be estimated at present, and may or may not be sufficient for practical expermental studies. An observation of the expected strange states $Z_{cs}$ would certainly significantly advance studies of hadrocharmonium.

I acknowledge illuminating discussions with Alexander Bondar.
This work is supported in part by U.S. Department of Energy Grant No.\ DE-SC0011842.

\end{document}